\documentclass[lettersize,journal]{IEEEtran}
\usepackage{amsmath,amsfonts}
\usepackage{algorithmic}
\usepackage{algorithm}
\usepackage{array}
\usepackage[caption=false,font=normalsize,labelfont=sf,textfont=sf]{subfig}
\usepackage{textcomp}
\usepackage{stfloats}
\usepackage{url}
\usepackage{verbatim}
\usepackage{graphicx}
\usepackage{multirow}
\usepackage{cite}
\IEEEpubidadjcol

\begin{document}

\title{An Open Multi-Center Whole-Body FDG PET/CT Foundation Model for Tumor Segmentation}

\author{Xiaofeng Liu,~\IEEEmembership{Member,~IEEE}, Qianru Zhang, Thibault Marin,~\IEEEmembership{Member,~IEEE}, Menghua Xia, Chi Liu,\\ Georges El Fakhri,~\IEEEmembership{Fellow,~IEEE,} Jinsong Ouyang,~\IEEEmembership{Senior Member,~IEEE}%

        % <-this % stops a space
\thanks{Xiaofeng Liu, Qianru Zhang, Thibault Marin, Menghua Xia, Chi Liu, Georges El Fakhri, and Jinsong Ouyang are with the Department of Radiology and Biomedical Imaging, Yale Biomedical Imaging Institute, Yale University, New Haven, CT, USA.}%

% <-this % stops a space
\thanks{Corresponding author: Jinsong Ouyang (email: jinsong.ouyang@yale.edu).}}

\markboth{Journal of \LaTeX\ Class Files,~Vol.~14, No.~8, August~2021}%
{Shell \MakeLowercase{\textit{et al.}}: A Sample Article Using IEEEtran.cls for IEEE Journals}

\IEEEpubid{0000--0000/00\$00.00~\copyright~2021 IEEE}

\maketitle

\begin{abstract}
The synergistic interpretation of anatomical information from computed tomography (CT) and metabolic information from positron emission tomography (PET) is important to oncologic imaging. However, existing deep learning methods for PET/CT remain largely task-specific, are often trained on single-center cohorts, or adopt dual-branch fusion schemes that delay cross-modal interaction and underutilize early spatial correspondence between PET and CT.
To address these limitations, we present an open-source, multi-center, whole-body FDG PET/CT foundation model utilizing 4,997 harmonized scans from four public datasets. Our framework employs hierarchical UNet-shaped backbones with early channel-wise concatenation, enabling anatomical and metabolic features to interact from the first embedding layer onward.  We further introduce a masked autoencoding objective based on zero-mean imputation, combined with a weighted global reconstruction loss. This design avoids non-physical intensity discontinuities at masked-region boundaries that arise from learnable mask tokens.
On downstream AutoPET lesion segmentation, the proposed models demonstrate strong label efficiency: with only 10\% of the labeled training data, they achieve performance comparable to models trained from scratch on the full dataset. Under extreme 5-shot linear probing, joint PET/CT pretraining also achieves higher Dice scores than separated-modality pretraining. This multi-center foundation model demonstrates label efficiency and cross-modality representation learning for PET/CT tumor segmentation. It provides a robust, open-source basis for advancing automated oncologic imaging, significantly reducing the need for large-scale manual annotations in clinical practice.
\end{abstract}


\section{Introduction}
\label{sec:intro}
The complementary use of positron emission tomography (PET) and computed tomography (CT) has become a mainstay of oncologic imaging because it jointly characterizes anatomy and metabolism, enabling lesion detection, staging, and treatment monitoring \cite{eyuboglu2021multi,vali2021snmmi}. In addition to visual interpretation, accurate tumor delineation is essential for quantifying disease burden, supporting staging, guiding treatment planning, and enabling longitudinal assessment of therapeutic response. In routine clinical practice, however, PET/CT interpretation remains labor-intensive and error-prone. Radiologists often need to review hundreds of slices per study while distinguishing malignant uptake from normal physiologic FDG activity and other confounding patterns. This process is further complicated by substantial inter-reader and intra-reader variability, as lesion identification and boundary delineation can vary across radiologists and even for the same reader at different times~\cite{liu2026aidriven}.

Recent advances in deep learning have substantially improved automated PET/CT analysis, particularly for lesion segmentation and detection \cite{zhong2023pet,dong2024head}. Despite this progress, most existing methods have limited clinical generalizability. A large fraction of prior work is designed for a single downstream task and trained on relatively small, single-center cohorts \cite{oh2025developing}. Such approaches depend heavily on scarce lesion-level annotations and often exhibit reduced robustness across institutions, scanner vendors, acquisition protocols, and patient populations \cite{wi2026delving,liu2022act,bakas2018identifying}. Moreover, models trained with a relatively small number of studies may experience substantial performance degradation when applied to out-of-distribution cases.

Foundation models trained on large-scale and diverse datasets, often comprising tens of thousands of medical imaging volumes, offer significant advantages over conventional task-specific models trained on limited cohorts of hundreds or fewer studies. By learning generalizable representations without being optimized for a single downstream task, such models can substantially reduce the amount of labeled data required for a given application. In addition, they tend to be less sensitive to domain shifts arising from inter-institutional variations, scanner vendors, and acquisition protocols, and are more robust when applied to out-of-distribution cases. From an optimization perspective, foundation models often enable faster and more stable training during downstream fine-tuning, and provide a flexible framework for modular domain adaptation across different clinical tasks and datasets. Early efforts focused mainly on 2D modalities such as chest X-rays \cite{huang2024chest}, whereas more recent work has shifted toward 3D volumetric data. Tak et al. \cite{tak2026generalizable} introduced BrainIAC, a UNet-shaped brain MRI foundation model, demonstrating that large-scale masked pre-training can substantially improve both full fine-tuning and few-shot linear probing. For dense voxel-wise downstream tasks, masked reconstruction objectives such as masked autoencoder (MAE) \cite{he2022masked} are often better aligned than contrastive approaches such as SimCLR \cite{chen2020simple}, because reconstruction encourages retention of fine spatial detail that contrastive invariance may suppress.

\IEEEpubidadjcol

This paradigm has recently been extended to PET/CT. Oh et al. \cite{oh2025developing,oh2025developingarxiv} proposed FratMAE, a PET/CT foundation model based on separate PET and CT encoders. While this approach leverages prior advances in single-modality representation learning, it treats PET and CT as isolated inputs and delays cross-modal interaction. In addition, its decoder reconstructs feature tokens rather than voxel intensities, limiting compatibility with dense voxel-level tasks such as lesion segmentation. The absence of hierarchical skip connections further constrains the model’s ability to recover high-resolution details critical for detecting small lesions.

Another major challenge in PET/CT representation learning is the scale and diversity of available data. Public PET/CT datasets remain small compared with natural-image datasets \cite{gatidis2022whole}, and prior PET/CT pre-training has largely relied on single-center cohorts \cite{oh2025developing}. This lack of diversity limits robustness to domain shifts arising from differences in scanners, acquisition protocols, and patient populations. To address this issue, we aggregate multiple public FDG PET/CT datasets and resampled them into a unified physical voxel space (Table~\ref{tab:dataset_scales}), mitigating cross-dataset heterogeneity while preserving anatomically and metabolically meaningful image characteristics.

Masked autoencoders \cite{he2022masked} have become a standard approach for visual pre-training, but their conventional implementation is not fully compatible with hierarchical encoder-decoder architectures widely used in medical imaging. In ViT-based MAE, masked patches are replaced with learnable tokens \cite{oh2025developing}, which can introduce artificial discontinuities. In UNet-like architectures, such discontinuities may destabilize convolutional processing in skip connections and degrade high-frequency structural information. This issue is particularly critical for PET/CT, where accurate delineation of small lesions depends on preserving fine spatial details.

To address these limitations, we introduce a multi-center whole-body FDG PET/CT foundation model pre-trained on 4,797 harmonized scans from four public datasets, with an additional 200 scans for leave-out evaluation. Instead of using separate modality-specific encoders, we adopt early channel-wise concatenation within hierarchical UNet-shaped backbones, enabling joint learning of anatomical and metabolic features from the first embedding layer onward. We further design a masked autoencoding objective tailored to encoder-decoder architectures by replacing learnable mask tokens with dual-channel randomized zero-mean imputation and optimizing a weighted global reconstruction loss. This design preserves stable skip-connection behavior while emphasizing masked-region recovery.

\begin{table*}[t!]
\centering
\caption{Dataset scale, spatial extent, and demographic summary. $\dag$ Only the FDG-PET subset of DeepPSMA is used; PSMA-PET scans are excluded. $*$ Eleven known anomalous SPADE samples containing only a single slice were excluded.}
\resizebox{1.0\linewidth}{!}{
\renewcommand{\arraystretch}{1.2}
\begin{tabular}{l|c|cc|cc|cc|c|l|c|c}
\hline
Dataset & Valid FDG PET/CT Scans & Z Min & Z Max & Y Min & Y Max & X Min & X Max & Site(s) & Scanner Vendors & Female & Age \\ \hline
AutoPET IV~\cite{gatidis2022whole} & 1014 & 200 & 661 & 122 & 235 & 179 & 251 & Germany & SIEMENS & 44.2\% & 59.7$\pm$15.9 \\
DeepPSMA$\dag$~\cite{jackson_2025_15281784} & 100 & 296 & 608 & 151 & 293 & 196 & 369 & Australia & GE/SIEMENS & 0\% & - \\
SPADE~\cite{eyuboglu2024dataset,eyuboglu2021multi} & 1126$*$ & 103 & 307 & 109 & 224 & 141 & 224 & USA & GE & 52.1\% & 59.4$\pm$9.3 \\
ViMedPET~\cite{nguyen2025toward} & 2757 & 207 & 551 & 104 & 245 & 167 & 256 & Vietnam & GE & 34.1\% & 57.8$\pm$13.7 \\ \hline
\end{tabular}
}
\label{tab:dataset_scales}
\end{table*}

To the best of our knowledge, this is the first open-source PET/CT foundation model\footnote{Code and pre-trained weights are available at \url{https://github.com/liu-xiaofeng/Foundation-Model-for-PET-CT.git}.} pre-trained on a large multi-center FDG PET/CT corpus. Our contributions are threefold: %First, we curate and harmonize a large multi-center 3D whole-body FDG PET/CT corpus of 4,997 scans, substantially expanding the scale and diversity of public PET/CT representation learning. Second, we propose a masked autoencoding framework with independent PET/CT masking, zero-mean imputation, and weighted global reconstruction, making MAE training compatible with hierarchical CNN and hybrid CNN-Transformer backbones. Third, we demonstrate that the resulting foundation model family, including SwinUNETR-v2 and nnUNet-v2 variants, achieves strong label efficiency and substantially outperforms separated-modality pre-training in both fine-tuning and extreme few-shot linear probing.

%Our contributions are threefold:

\noindent$\bullet$ We curate and harmonize a large multi-center 3D whole-body FDG PET/CT pre-training corpus of 4,997 scans, substantially expanding the scale and diversity of public PET/CT representation learning.

\noindent$\bullet$ We propose a masked autoencoding framework with independent PET/CT masking, zero-mean imputation, and weighted global reconstruction, making MAE training compatible with hierarchical CNN and hybrid CNN-Transformer backbones.

\noindent$\bullet$ We demonstrate that the resulting foundation model family, including SwinUNETR-v2 and nnUNet-v2 variants, achieves strong label efficiency and substantially outperforms separated-modality pre-training in both fine-tuning and extreme few-shot linear probing.

\begin{figure}[!t]
\begin{center}
\includegraphics[width=1\linewidth]{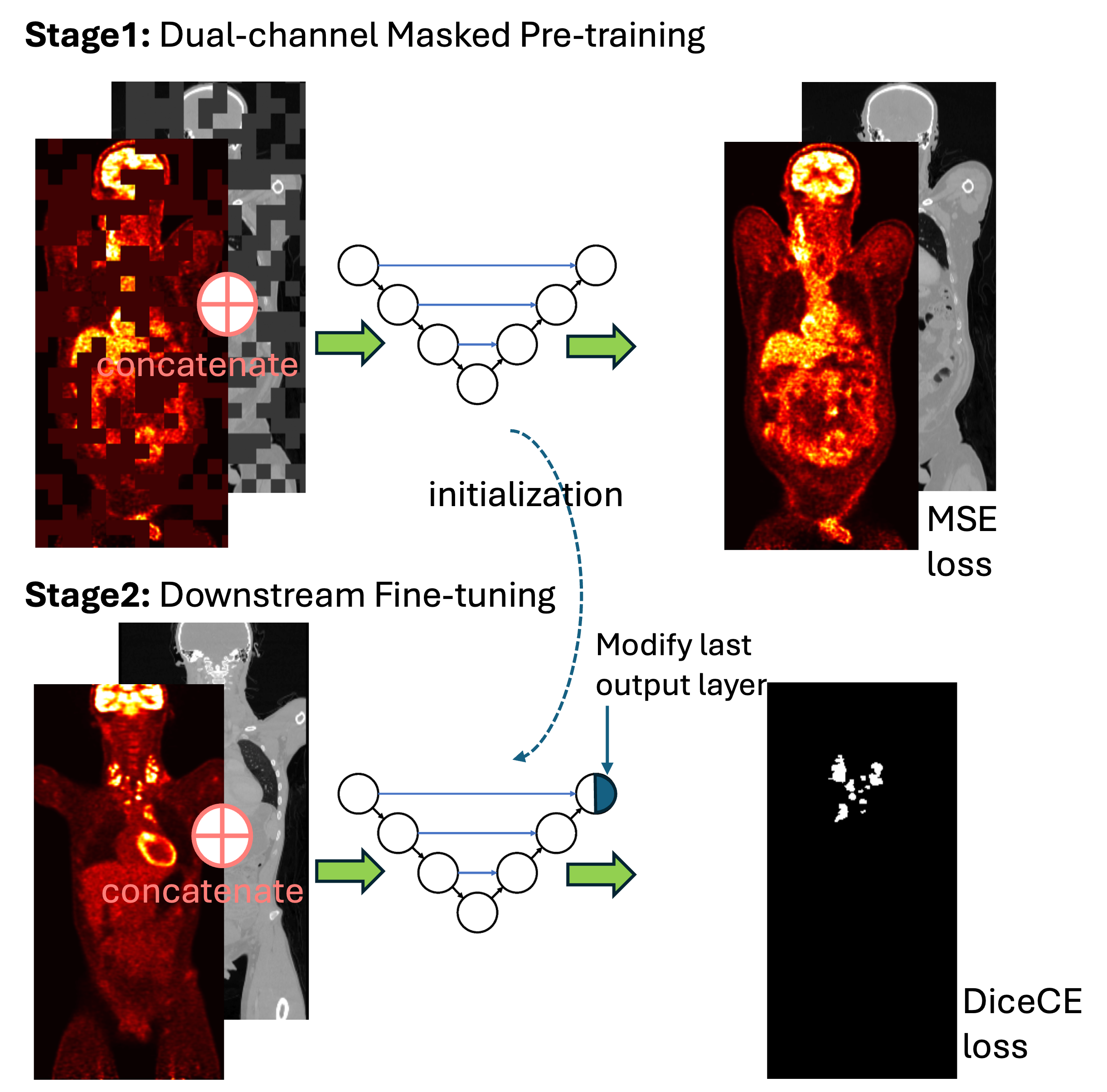}
\end{center}%%\vspace{-10pt}
\caption{Overview of the proposed dual-channel masked autoencoder for pre-training and downstream fine-tuning. PET and CT are masked independently.}
\label{fig:framework}
\end{figure}

\section{Materials and Methods}

\subsection{Multi-Center Data Curation}
Training a PET/CT foundation model that generalizes across institutions requires careful handling of inter-dataset heterogeneity. Existing PET/CT pre-training studies \cite{oh2025developing} are often restricted to a single center, limiting diversity in anatomy, physiologic uptake, acquisition protocols, and scanner characteristics. As summarized in Table~\ref{tab:dataset_scales}, our corpus contains 4,997 valid whole-body FDG PET/CT scans aggregated from four public datasets: ViMedPET, AutoPET IV, DeepPSMA, and SPADE.

We design a harmonization pipeline to standardize these data while preserving physical image meaning. For all datasets, preprocessing includes CT blank-boundary cropping, resampling to a unified voxel spacing of $2.0\times2.0\times3.0\,\mathrm{mm}^3$, and volume-wise $z$-score normalization performed separately for CT and PET. For ViMedPET, misregistration between PET and CT was observed for some subjects, likely due to patient motion during multi–bed-position PET acquisition. To address this, we applied mutual-information-based rigid PET-to-CT registration\footnote{\url{https://simpleitk.readthedocs.io/en/master/registrationOverview.html}}.

\begin{figure*}[t!]
\begin{center}
\includegraphics[width=1\linewidth]{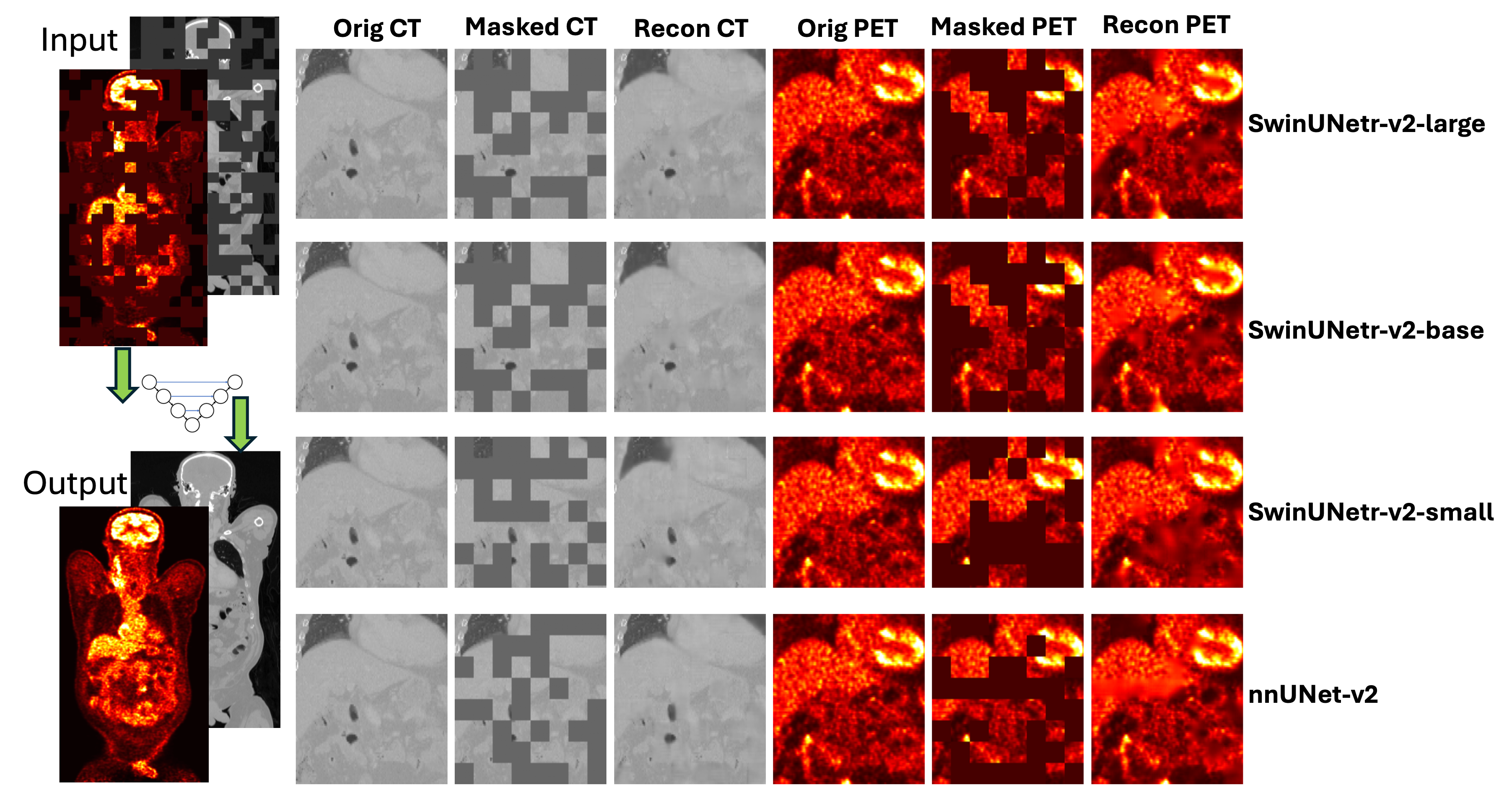}
\end{center}\vspace{-10pt}
\caption{Examples of masked CT and PET patches and the corresponding reconstructions using SwinUNETR-v2-large/base/small and nnUNet-v2-3d\_fullres as the foundation-model backbones.}
\label{fig:reconstructed_results}
\end{figure*}

A key design choice is to standardize images in physical space rather than resizing them to identical matrix sizes. Naively forcing all scans to a fixed array shape, as in some prior work \cite{oh2025developing}, can distort anatomy and alter the physical scale of small lesions. Instead, we resample each scan to a common voxel spacing while allowing variable matrix sizes. This preserves anatomical geometry and lesion scale and encourages the model to learn physically meaningful priors across datasets.

To standardize CT Hounsfield Units and PET standardized uptake values (SUVs), we apply volume-wise $z$-score normalization:
\begin{equation}
 x_{\mathrm{norm}} = \frac{x - \mu}{\max(\sigma, 1 \times 10^{-8})}.
\end{equation}
The normalized CT and PET volumes are then concatenated channel-wise to form the input tensor $x \in \mathbb{R}^{2 \times Z \times Y \times X}$. Unlike dual-branch architectures \cite{oh2025developing}, this design ensures that the first embedding layer already receives coupled anatomical and metabolic information.

\subsection{Multimodal Backbone Architecture}
We build our foundation models on encoder-decoder architectures that are already strong performers in medical image segmentation. Specifically, we evaluate SwinUNETR-v2 \cite{he2023swinunetr} at two scales and nnUNet-v2 \cite{isensee2021nnu}. During pre-training, each input crop has spatial size $96 \times 128 \times 128$ in the $Z$, $Y$, and $X$ dimensions. This anisotropic crop size is matched to the standardized voxel spacing of $2.0\times2.0\times3.0\,\mathrm{mm}^3$, giving the model a more balanced physical field of view across axes.

SwinUNETR \cite{hatamizadeh2021swin,he2023swinunetr} uses a hierarchical shifted-window Transformer encoder with multi-scale skip connections into a convolutional decoder. This architecture is well-suited to PET/CT because it combines long-range contextual modeling with high-resolution structural recovery. We use the MONAI default setting\footnote{\url{https://monai-dev.readthedocs.io/en/fixes-sphinx/_modules/monai/networks/nets/swin_unetr.html#SwinUNETR}}, feature size $=24$ and depths $(2,2,2,2)$, as SwinUNETR-v2-small, and the challenge-winning setting, feature size $=48$ and depths $(2,2,6,2)$, as SwinUNETR-v2-base. Moreover, we also trained a large version with feature size $=96$ and depth $(2,2,18,2)$ following the SwinTransformer~\cite{liu2021swin} setting.

\begin{table}[t!]
\centering
\caption{Backbone model size comparison in terms of total parameter count and model file size.}
\resizebox{1\columnwidth}{!}{%
\renewcommand{\arraystretch}{1}
\begin{tabular}{lcc}
\hline
Model & Total Params & Size (MB) \\
\hline
SwinUNETRv2-large (2,2,18,2)-96~\cite{he2023swinunetr} & 319.208M & 1217.68 \\
SwinUNETRv2-base (2,2,6,2)-48~\cite{he2023swinunetr} & 74.649M & 284.76 \\
SwinUNETRv2-small (2,2,2,2)-24~\cite{he2023swinunetr} & 18.348M & 69.99 \\
nnUNetV2-3d\_fullres PlainConvUNet~\cite{isensee2021nnu} & 31.196M & 119.01 \\
\hline
\end{tabular}}
\label{tab:model_size}
\end{table}

We also include nnUNet-v2, which has been regarded as a strong convolutional baseline in medical image segmentation. Specifically, we leverage its \textit{3d\_fullres} configuration equipped with the \textit{PlainConvUNet} architecture. Unlike aggressive downsampling strategies, the \textit{3d\_fullres} pipeline processes the concatenated PET/CT volumes at their native unified voxel spacing, thereby preventing the spatial degradation of small lesions. This model serves as a purely convolutional counterpart to the Transformer-based SwinUNETR family.

All of these backbones are naturally suited to downstream segmentation because both the encoder and decoder are retained. Only the final projection layer needs to be changed for the downstream target labels. Note that the encoder can be used to extract features for broader downstream classification and regression tasks, as in~\cite{tak2026generalizable}.

\begin{figure*}[!t]
\begin{center}
\includegraphics[width=1\linewidth]{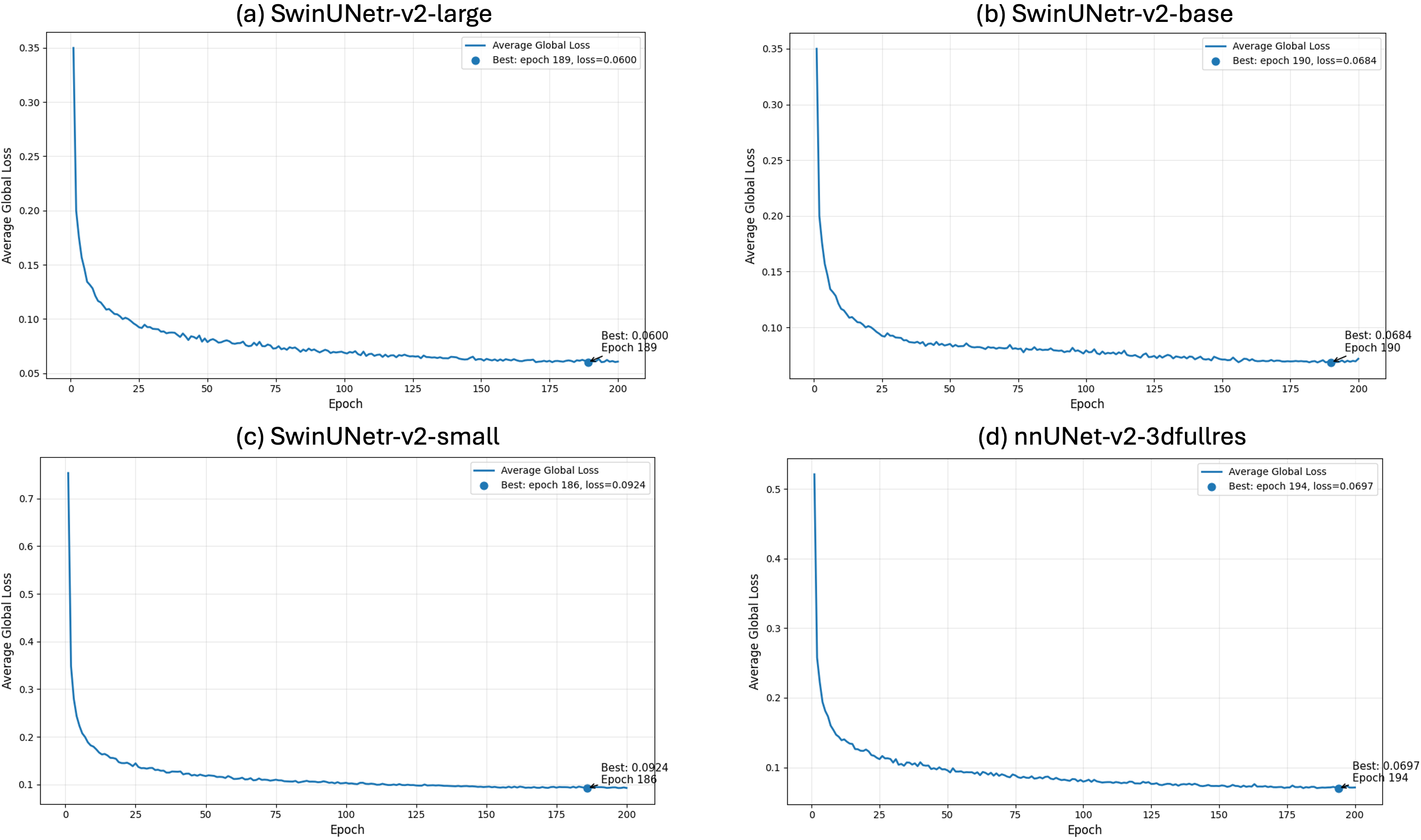}
\end{center}\vspace{-10pt}
\caption{Training loss curves for the three foundation-model backbones: (a) SwinUNETR-v2-large, (b) SwinUNETR-v2-base, (c) SwinUNETR-v2-small, and (d) nnUNet-v2-3d\_fullres.}
\label{fig:loss_curves}
\end{figure*}

\subsection{Zero-Mean-Imputation MAE}
To encourage cross-modal reasoning while remaining compatible with encoder-decoder backbones, we formulate a masked autoencoding task directly in PET/CT image space. As shown in Figure~\ref{fig:reconstructed_results}, the input volume $x$ is partitioned into non-overlapping 3D patches of size $12\times16\times16$ along the $Z$, $Y$, and $X$ axes. We define a binary mask $M \in \{0,1\}^{B\times2\times Z\times Y\times X}$, where 1 indicates a masked voxel, and 0 indicates a visible voxel. PET and CT are masked independently, and we use a masking ratio of 50\%.

Instead of utilizing a learnable mask token, which injects artificial constant values into the feature maps and disrupts the 3D convolutional layers in SwinUNETR-v2 and nnUNet models, we exploit our Z-score normalization. Since the expected value of the background and normalized intensities is approximately zero, we perform zero-mean imputation:
\begin{equation}
\tilde{x} = x \odot (1 - M) + 0 \odot M = x \odot (1 - M),
\end{equation}
where $\odot$ denotes element-wise multiplication. The masked volume $\tilde{x}$ is then passed through the backbone to reconstruct the original multi-modal volume, denoted by $x'_{rec} = f(\tilde{x})$.

To intuitively demonstrate the efficacy of this strategy, Figure~\ref{fig:reconstructed_results} shows representative masked inputs and reconstructions for SwinUNETR-v2-based large/base/small models and the nnUNet-v2-3d\_fullres model. In all cases, the model reconstructs visually plausible content. When only PET or CT is masked, the network often recovers the missing content well by exploiting cross-modal correlations. When both modalities are masked in the same region, reconstructions are naturally blurrier because local information is absent. This is expected, and exact photorealistic reconstruction is not the goal; rather, the objective is to learn transferable joint representations.

\subsection{Weighted Global Reconstruction Loss}
Standard MAE optimizes mean squared error only on the masked region. For hierarchical UNet-shaped architectures, however, completely ignoring visible regions can destabilize training because skip pathways no longer receive pressure to preserve consistent identity information. To address this issue, we use a weighted global reconstruction loss. Let
\begin{equation}
\mathcal{L}_{\mathrm{MSE}} = (x'_{rec} - x)^2,
\end{equation}
be the voxel-wise squared reconstruction error. We define the total loss as
\begin{align}
\mathcal{L}_{\mathrm{total}} =
\frac{\sum (\mathcal{L}_{\mathrm{MSE}} \odot M)}{\sum M + \epsilon}
+ \lambda \frac{\sum (\mathcal{L}_{\mathrm{MSE}} \odot (1 - M))}{\sum (1 - M) + \epsilon},
\end{align}
where $\lambda=0.2$ in all experiments based on grid searching from $\{0.1,0.2,0.25,0.5\}$. The first term emphasizes reconstruction of masked regions, while the second provides a weaker constraint on visible regions to stabilize decoder and skip-connection learning.

\begin{figure*}[!t]
\begin{center}
\includegraphics[width=1\linewidth]{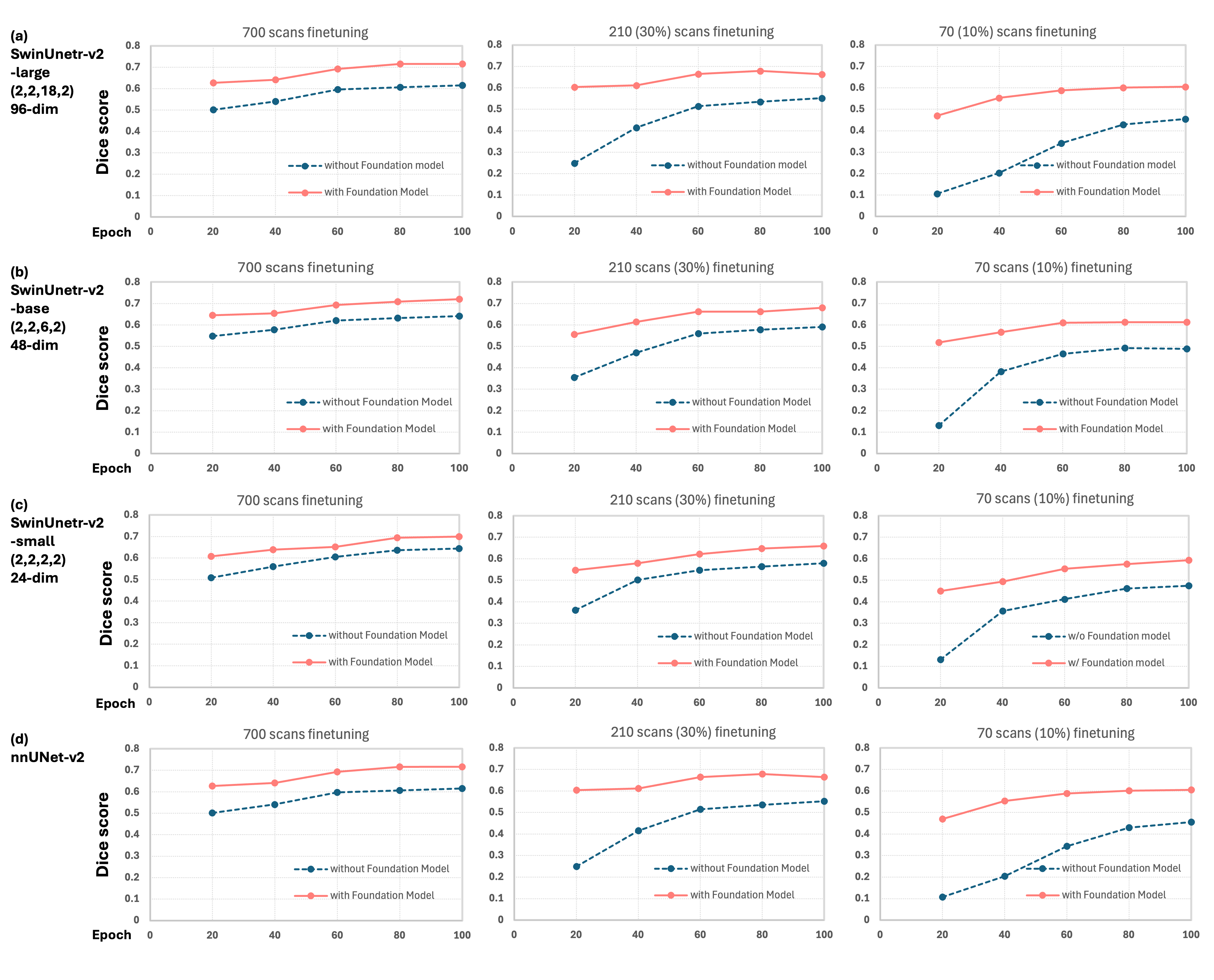}
\end{center}\vspace{-5pt}
\caption{Dice score on downstream AutoPET lesion segmentation under different fine-tuning data scales and backbone choices, comparing training from scratch with MAE initialization.}
\label{fig:dice_comparison}
\end{figure*}

\section{Results}
We evaluate the proposed foundation models on downstream AutoPET lesion segmentation. To avoid information leakage, we hold out 200 subject-independent AutoPET scans for testing. The remaining 814 AutoPET scans, together with the other three datasets, are used for pre-training, yielding 4,797 scans in total for self-supervised learning in MAE pre-training (stage 1). All foundation models are pre-trained for 200 epochs. Figure~\ref{fig:loss_curves} shows stable convergence across different backbones. Pre-training takes approximately 20 to 25 hours on two NVIDIA A100 80GB GPUs using batch size 16 and an initial learning rate of $1\times10^{-4}$ for SwinUNETR or $5\times10^{-4}$ for nnUNet with cosine annealing following previous MRI-only models~\cite{tak2026generalizable}. Whole-body inference uses sliding-window evaluation.

For downstream segmentation, the 814 remaining AutoPET scans are split into 700 training and 114 validation scans, while the 200 held-out scans are used only for testing. The target is to segment hypermetabolic tumor lesions from background physiologic uptake. Fine-tuning uses Dice plus cross-entropy loss (DiceCE) \cite{tak2026generalizable}.

\begin{table*}[t!]
\centering
\setlength{\tabcolsep}{4pt}
\caption{Full-parameter fine-tuning results for downstream AutoPET lesion segmentation under different backbones and different proportions of labeled downstream training data. All MAE trained with 4,797 scans.  Best results are in bold and second-best results are underlined.}%\vspace{-5pt}
\renewcommand{\arraystretch}{1.2}
\resizebox{1\textwidth}{!}{%
\begin{tabular}{l|cc|cc|cc|cc|cc|cc}
\hline
\multirow{3}{*}{Model}
& \multicolumn{4}{c|}{700 scans (100\%) fine-tuning}
& \multicolumn{4}{c|}{210 scans (30\%) fine-tuning}
& \multicolumn{4}{c}{70 scans (10\%) fine-tuning} \\
\cline{2-13}
& \multicolumn{2}{c|}{Dice $\uparrow$} & \multicolumn{2}{c|}{HD95 $\downarrow$}
& \multicolumn{2}{c|}{Dice $\uparrow$} & \multicolumn{2}{c|}{HD95 $\downarrow$}
& \multicolumn{2}{c|}{Dice $\uparrow$} & \multicolumn{2}{c}{HD95 $\downarrow$} \\
\cline{2-13}
& Scratch & MAE & Scratch & MAE
& Scratch & MAE & Scratch & MAE
& Scratch & MAE & Scratch & MAE \\
\hline
SwinUNETR(v2)-large
& 0.6296 &  {0.7047} & 99.5634 &  {64.3318}
& 0.5514 & {0.6541} & 110.9445 &  {68.8843}
& 0.4715 & {0.5782} & 112.7857 &  {79.5413} \\

SwinUNETR(v2)-base
& 0.6414 & \textbf{0.7205} & 94.3842 & \textbf{59.1572}
& 0.5904 & \textbf{0.6801} & 101.4528 & \textbf{63.4412}
& 0.4889 & \textbf{0.6132} & 113.3982 & \textbf{70.3947} \\

SwinUNETR(v2)-small
& 0.6440 & 0.6991 & 95.3705 & 70.5946
& 0.5793 & 0.6594 & 101.3859 & 80.0549
& 0.4751 & 0.5930 & 106.1070 & 84.5757 \\

nnUNet(v2)-3d\_fullres
& 0.6149 & \underline{0.7160} & 85.3512 & \underline{63.5110}
& 0.5525 & \underline{0.6642} & 88.7968 & \underline{68.6980}
& 0.4555 & \underline{0.6055} & 116.2797 & \underline{77.2990} \\
\hline
\end{tabular}}
\label{tab:full_finetuning}
\end{table*}

\subsection{Fine-Tuning Under Data Scarcity}
Following the data-scaling protocol of BrainIAC \cite{tak2026generalizable}, we create three downstream training subsets containing 10\% (70 scans), 30\% (210 scans), and 100\% (700 scans) of the AutoPET training set. We compare MAE-initialized models with randomly initialized baselines across SwinUNETR-v2-large, SwinUNETR-v2-base, SwinUNETR-v2-small, and nnUNet-v2. All models are fine-tuned for 100 epochs using an initial learning rate of $1\times10^{-4}$ with cosine annealing. With a batch size of 4, the SwinUNETR-base takes about 22G GPU memory, while nnUNet-v2 takes about 10G GPU memory.

The results in Table~\ref{tab:full_finetuning} and Figure~\ref{fig:dice_comparison} show a clear and consistent advantage for MAE initialization across all architectures and data regimes. Even with 700 training scans, the improvement of the Dice score is over 5\% for all backbones. Notably, the SwinUNETR-v2-large is able to achieve the lowest loss, though this metric is not the main objective of pre-training. We can see that its segmentation performance is consistently inferior to SwinUNETR-v2-base. Under limited downstream data, the large model is more prone to overfitting and struggles to effectively map its representations to dense segmentation predictions.

\begin{figure*}[t!]
\begin{center}
\includegraphics[width=1\linewidth]{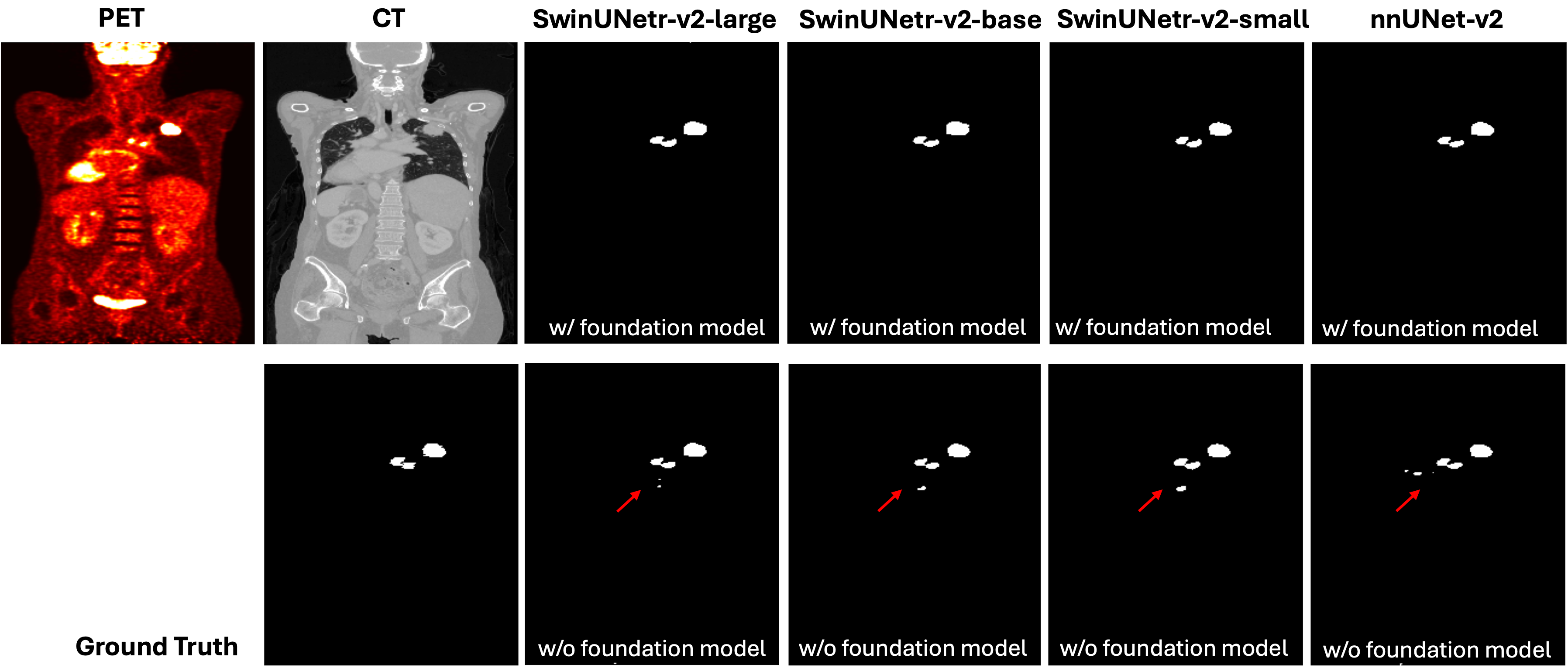}
\end{center}\vspace{-5pt}
\caption{Qualitative comparison of downstream AutoPET lesion segmentation with different foundation-model backbones.}
\label{fig:qualitative_results}
\end{figure*}

To visually corroborate our quantitative findings, Figure~\ref{fig:qualitative_results} presents a qualitative comparison of downstream AutoPET lesion segmentation under the extreme $100\%$ data regime. MAE-initialized models are able to generate lesion masks that are contiguous, anatomically plausible, and aligned with the reference labels. Note that with large-scale data to support the pre-training, the SwinUNETR-v2-base usually demonstrates the best performance among those backbones. It exhibits fewer false negatives (top example) and false positives (bottom example), respectively. The visual fidelity of our model is directly attributable to the robust pixel-level reconstruction capability learned during the zero-mean imputation MAE phase, as previously evidenced by the high-fidelity patch reconstructions in Figure~\ref{fig:reconstructed_results}.

MAE initialization consistently yielded substantial performance gains over scratch training, particularly in the 10\% data regime (Table 3). Importantly, the MAE-initialized model trained with only 70 labeled scans performs competitively with the model trained from scratch with all 700 scans. This gap strongly suggests that large-scale multi-center pre-training provides transferable anatomical and metabolic priors that reduce downstream annotation demands.

\subsection{Ablation on Pre-Training Regime and Modality Fusion}
We next isolate the effect of pre-training corpus scale and modality-fusion strategy. Results are shown in Table~\ref{tab:ablation_regimes}.

First, restricting pre-training to the single-center AutoPET subset of 814 scans consistently reduces downstream performance relative to pre-training on the full 4,797-scan multi-center corpus. For example, at 100\% downstream fine-tuning, Dice drops from $0.7205$ to $0.6683$. This result indicates that broader multi-center pre-training improves representation robustness and reduces domain overfitting.

Second, we compare our early-concatenation design with a separated-modality encoders scheme that conceptually mirrors dual-encoder PET/CT pre-training \cite{oh2025developing}. Across all data regimes, early concatenation performs better. This suggests that forcing PET and CT to interact from the input stage yields a more useful latent space than learning the two modalities in parallel and fusing them later.

\begin{table*}[t!]
\centering
\setlength{\tabcolsep}{4pt}
\caption{Full-parameter fine-tuning ablation on pre-training data regime and modality-fusion strategy for downstream AutoPET lesion segmentation using SwinUNETR. ``-'' indicates that scratch initialization is unaffected by the pre-training regime. Best results are in bold, and second-best results are underlined.}%\vspace{-5pt}
\renewcommand{\arraystretch}{1.2}
\resizebox{1\textwidth}{!}{%
\begin{tabular}{l|cc|cc|cc|cc|cc|cc}
\hline
\multirow{3}{*}{Model}
& \multicolumn{4}{c|}{700 scans (100\%) fine-tuning}
& \multicolumn{4}{c|}{210 scans (30\%) fine-tuning}
& \multicolumn{4}{c}{70 scans (10\%) fine-tuning} \\
\cline{2-13}
& \multicolumn{2}{c|}{Dice $\uparrow$} & \multicolumn{2}{c|}{HD95 $\downarrow$}
& \multicolumn{2}{c|}{Dice $\uparrow$} & \multicolumn{2}{c|}{HD95 $\downarrow$}
& \multicolumn{2}{c|}{Dice $\uparrow$} & \multicolumn{2}{c}{HD95 $\downarrow$} \\
\cline{2-13}
& Scratch & MAE & Scratch & MAE
& Scratch & MAE & Scratch & MAE
& Scratch & MAE & Scratch & MAE \\
\hline
SwinUNETR(v2)-base [4,797]
& \textbf{0.6414} & \textbf{0.7205} & \textbf{94.3842} & \textbf{59.1572}
& \textbf{0.5904} & \textbf{0.6801} & \textbf{101.4528} & \textbf{63.4412}
& \textbf{0.4889} & \textbf{0.6132} & \textbf{113.3982} & \textbf{70.3947} \\

SwinUNETR(v2)-base [814 AutoPET]
& - & 0.6683 & - & \underline{72.0771}
& - & 0.6337 & - & \underline{82.3765}
& - & \underline{0.5788} & - & \underline{86.7747} \\
\hline
Separate PET/CT [4,797]~\cite{oh2025developing}
& 0.6124 & \underline{0.6837} & 97.6594 & 73.6302
& 0.5584 & \underline{0.6409} & 104.5289 & 84.0576
& 0.4594 & 0.5663 & 108.4414 & 87.8742 \\
\hline
\end{tabular}}
\label{tab:ablation_regimes}
\end{table*}

\subsection{Few-Shot Linear Probing}
To evaluate the intrinsic quality of the learned features without substantial downstream adaptation, we conduct extreme few-shot linear probing following \cite{tak2026generalizable}. We freeze the encoder and most of the decoder and update only the final decoder layer using $K=5$ labeled scans. Following standard practice for linear probing, we use a higher initial learning rate of $1\times10^{-3}$ with cosine annealing.

Table~\ref{tab:few_shot_linear} shows that training from scratch nearly collapses under this protocol, with Dice scores around 0.08 to 0.09. In contrast, MAE pre-training yields clearly usable representations. Moreover, joint PET/CT pre-training substantially outperforms separated-modality pre-training, for example, improving Dice from $0.2163$ to $0.3557$ for SwinUNETR(v2)-base. This result supports the claim that early fusion improves not only fine-tuning performance but also the underlying structure of the learned latent space.

Notably, the linear probing results demonstrate that the large-capacity model successfully avoids representation collapse. Aligning with its pre-training reconstruction loss, SwinUNETR-v2-large achieves the highest linear probing performance among the variants ($0.3561$ vs. $0.3557$ Dice for the base model). This confirms that its frozen latent space is highly linearly separable, effectively capturing robust, high-level semantic priors ready for downstream adaptation.

\begin{table*}[t]
\centering
\setlength{\tabcolsep}{5pt}
\caption{Few-shot $K=5$ linear-probing (only last layer is fine-tuned) results on downstream AutoPET lesion segmentation. Best results are in bold and second-best results are underlined.}%\vspace{-5pt}
\renewcommand{\arraystretch}{1.2}
\resizebox{0.7\textwidth}{!}{%
\begin{tabular}{l|cc|cc|cc}
\hline
\multirow{2}{*}{Model}
& \multicolumn{2}{c|}{Scratch}
& \multicolumn{2}{c|}{MAE-Separate-PET/CT as~\cite{oh2025developing}}
& \multicolumn{2}{c}{MAE-Concat-PET/CT} \\
\cline{2-7}
& Dice $\uparrow$ & HD95 $\downarrow$ & Dice $\uparrow$ & HD95 $\downarrow$ & Dice $\uparrow$ & HD95 $\downarrow$\\
\hline
SwinUNETR(v2)-large
& 0.0674 & 135.632 & \underline{0.1475} & \underline{116.760} & \textbf{0.3561} & \textbf{96.898} \\

SwinUNETR(v2)-base
& 0.0876 & 129.4734 & \underline{0.2163} & \underline{108.5311} & \textbf{0.3557} & \textbf{97.4730} \\

SwinUNETR(v2)-small
& 0.0825 & 125.4951 & \underline{0.1010} & \underline{111.6708} & \textbf{0.3326} & \textbf{102.935} \\

nnUNet(v2)-3d\_fullres
& 0.0911 & 134.6174 & \underline{0.1959} & \underline{108.7083} & \textbf{0.3497} & \textbf{95.5463} \\
\hline
\end{tabular}}
\label{tab:few_shot_linear}
\end{table*}

\section{Discussion}
\label{sec:discussion}

We curated 4,997 harmonized multi-center scans and developed an open-source, pre-trained whole-body FDG PET/CT foundation model family. By standardizing images in physical voxel space rather than enforcing a shared matrix size, the proposed pipeline preserves lesion scale and anatomical geometry while reducing inter-dataset heterogeneity.

Methodologically, our framework differs from recent PET/CT foundation-model designs in two key aspects. First, it employs early PET/CT concatenation within hierarchical encoder–decoder backbones, enabling joint anatomical–metabolic representation learning from the earliest layers. Second, it replaces conventional mask tokens with zero-mean imputation and introduces a weighted global reconstruction loss that is compatible with hybrid CNN–Transformer UNet-shaped architectures. Together, these design choices make masked autoencoding effective for high-resolution 3D medical imaging backbones used in downstream segmentation.

Across fine-tuning and few-shot linear probing experiments, the proposed models demonstrate strong transferability, label efficiency, and robustness. In particular, large-scale multi-center pre-training enables competitive downstream segmentation performance even when only 10\% of labeled AutoPET data are available. These findings indicate that multi-center PET/CT foundation models provide a strong and practical initialization for a broad range of oncologic imaging applications.

An important observation from our study is the scaling behavior of 3D vision backbones under downstream data constraints, which directly impacts clinical translation. Although the SwinUNETR-v2-large variant achieves the lowest MAE reconstruction loss and exhibits highly competitive few-shot linear probing performance, its full-parameter fine-tuning performance lagged behind the base variant. This discrepancy highlights a downstream data bottleneck: while our 4,797-scan corpus successfully instills robust, linearly separable representations into large-capacity models, the downstream cohort of 700 scans remains insufficient for adapting such massive parameter spaces, making the model prone to overfitting. Consequently, intermediate-capacity backbones like SwinUNETR-v2-base provide a better empirical balance between representational power and available annotated data. Beyond label efficiency, this allows the architecture to be flexibly tailored to specific institutional hardware constraints. In resource-constrained clinical environments ($\le$ 24GB VRAM), the nnUNet-v2 framework or the SwinUNETR-small variant offers a practical, memory-efficient solution. Conversely, for institutions with high-end accelerators ($\ge$ 48GB VRAM) and sufficient labeled data, fine-tuning the SwinUNETR-base maximizes the utility of the pre-trained multi-modal priors.

Despite these promising results, several limitations remain. Conventional pixel-level metrics (e.g., Dice) may not fully reflect clinically relevant lesion-level detectability and diagnostic uncertainty~\cite{pangrethinking,liu2026aidriven}. In addition, the current framework assumes the availability of both PET and CT modalities, whereas real-world scenarios may involve missing or incomplete modalities. Extending the model to handle such settings is an important direction for future work~\cite{zhaounimrseg,li2025deep}. Furthermore, integrating the proposed vision backbone with multimodal large language models~\cite{bai2024m3d} and radiology reports could enable weakly supervised lesion grounding and more interpretable, report-oriented PET/CT systems.

In the SPADE and ViMedPET datasets, intensity discontinuities are observed at certain axial locations, often near bed transition regions. These effects are likely associated with the multi–bed-position acquisition and reconstruction process in whole-body PET imaging, although the exact cause cannot be definitively established from the available data. Notably, SPADE and ViMedPET constitute a substantial portion of the multi-center corpus and reflect realistic clinical acquisition conditions. While such discontinuities are not ideal from a strictly homogeneous data perspective, retaining these datasets preserves the scale and diversity necessary for learning robust and generalizable representations. In practice, exposure to these variations may enhance robustness to real-world data heterogeneity. Although minimizing such effects may be beneficial in a fully controlled setting with standardized acquisition and reconstruction protocols, incorporating data that reflect typical acquisition variability is important for developing models that generalize across institutions. Future work may further investigate artifact-aware preprocessing or physics-informed correction strategies to mitigate these effects.

%\hl{Another practical consideration is the presence of acquisition-related artifacts in the pre-training data. In the SPADE and ViMedPET datasets, whole-body PET images are reconstructed by stitching multiple bed positions acquired sequentially along the axial direction. This process can introduce boundary artifacts at bed transition regions due to variations in count statistics, attenuation and scatter correction inconsistencies, and imperfect normalization across bed overlaps. These artifacts may manifest as intensity discontinuities or subtle structural inconsistencies, which can affect both visual interpretation and quantitative analysis.}

%\hl{Notably, SPADE and ViMedPET constitute a substantial portion of the pre-training corpus. While these artifacts are not ideal for representation learning, excluding these datasets would significantly reduce data scale and diversity, which are critical for training robust foundation models. Consequently, the model is exposed to both true anatomical and metabolic patterns as well as acquisition-induced variations, and may partially allocate representational capacity to modeling these non-physiological features. We anticipate that improved artifact mitigation, through artifact-aware preprocessing or physics-informed correction, could further enhance representation quality and downstream performance.}

\section{Conclusion}
\label{sec:conclusion}

In this work, we present a multi-center, whole-body FDG PET/CT foundation model pre-trained on 4,797 harmonized scans. By combining early cross-modal fusion with a zero-mean-imputation masked autoencoding strategy and a weighted global reconstruction loss, the proposed framework enables effective representation learning for high-resolution 3D medical imaging. Extensive experiments demonstrate that the resulting models achieve strong transferability, improved robustness, and substantial label efficiency, maintaining competitive performance even under limited labeled data regimes. These results highlight the potential of large-scale, multi-center PET/CT foundation models as a general-purpose initialization for a wide range of oncologic imaging tasks. Overall, this work provides a practical and scalable foundation for future developments in multimodal medical image analysis, including extensions to missing-modality learning, multimodal reasoning with language models, and clinically oriented decision support systems.

\section*{Disclosure}
The authors declare that they have no relevant financial or non-financial interests to disclose.

\section{Ethical Responsibilities}\vspace{-3pt}
This retrospective study used open-access human subject data; no additional ethical approval was required.

\section*{Acknowledgments} 
This work is supported in part by the NIH grants P41EB022544, R01CA275188, R21EB034911, R01CA296305, R01CA290745, and NVIDIA Academic Grant Program.

{
    %\small
    %\bibliographystyle{ieeenat_fullname}
    %\bibliography{main}
    \bibliographystyle{IEEEtran}
    \bibliography{main}
}

\end{document}